\documentclass[prc,twocolumn,showpacs,preprintnumbers,amsmath,amssymb,superscriptaddress,floatfix,nofootinbib]{revtex4}
\usepackage{graphicx}
\usepackage{amsmath}
\usepackage{amsfonts}
\usepackage{slashed}
\usepackage{amssymb}
\usepackage{hyperref}
\hypersetup{breaklinks=true}
\newcommand{\sm}[1]{\scriptscriptstyle{#1}}

\newcommand{\boldm}[1]{\mbox{\boldmath{$#1$}}}
  \def\CL{{\cal L}}
\def\CM{{\cal M}}

\begin{document}

\title{Search for the $\Sigma(1380)1/2^-$ state in $\Lambda^+_c \to \gamma \pi^+ \Lambda$ decay by triangle singularity}

\author{Ke Wang}
\affiliation{School of Nuclear Science and Technology, University of Chinese Academy of Sciences, Beijing 101408, China}

\author{Yu-Fei Wang}
\affiliation{School of Nuclear Science and Technology, University of Chinese Academy of Sciences, Beijing 101408, China}

\author{Bo-Chao Liu}
\affiliation{Ministry of Education Key Laboratory for Nonequilibrium Synthesis and Modulation of Condensed Matter, School of Physics, Xi'an Jiaotong University, Xi'an 710049, China}

\author{Fei Huang}
\email{huangfei@ucas.ac.cn}
\affiliation{School of Nuclear Science and Technology, University of Chinese Academy of Sciences, Beijing 101408, China}

\date{\today}

\begin{abstract}
In this work, we investigate the resonance production in $\Lambda^+_c \to \gamma \pi^+ \Lambda$ decay through the triangle singularity (TS) mechanism, within an effective Lagrangian approach. We find that the appropriate loop decay process could develop a triangle singularity in the invariant mass $M_{\pi\Lambda}$ around $1.35$ GeV, with the shape depending on the quantum numbers of $\Sigma^*$ states that couple to the final $\pi\Lambda$ system. Especially, the $\Sigma(1380)1/2^-$ state newly predicted in the pentaquark model, if exists, significantly enhances the TS contribution and sharpens the TS peak due to the $S$-wave $\Sigma^*\pi\Sigma$ vertex in the loop. Therefore, the existence of the $\Sigma(1380)1/2^-$ state can be examined by measuring the TS signal in $\Lambda^+_c \to \gamma \pi^+ \Lambda$. Considering also possible tree-level diagrams, we additionally obtain the branching ratio ${\rm Br}(\Lambda^+_c \to \gamma \pi^+ \Lambda)\simeq 1.5\times 10^{-5}$. We suggest the investigation of this reaction by future BESIII, LHCb, and Belle experiments.
\end{abstract}

\maketitle

\section{INTRODUCTION}
\label{introductiuon}

The study of the internal constituents of baryons is one of the important topics in hadron physics. Conventional quark models assume that a baryon consists of three constituent quarks and successfully describe the properties of the baryon spatial ground states, but encounter difficulties in explaining the long-standing mass reverse problem for the lowest spatially-excited baryon with spin-parity $J^P=1/2^-$~\cite{Zhang:2004xt,Helminen:2000jb}. Pentaquark models provide possible solutions for this problem by pulling an additional $q\bar{q}$ pair from the gluon field and simultaneously predict a possible baryon state $\Sigma^*(J^P=1/2^-)$ around $1360$ MeV~\cite{Zhang:2004xt} or $1405$ MeV~\cite{Helminen:2000jb}. Especially, in recent years, multi-quark models have made significant progress in shedding light on the properties of heavy hadrons, such as the $X(3872)$~\cite{Kalashnikova:2018vkv}, $P_c$~\cite{LHCb:2019kea}, and $T_{cc}$~\cite{LHCb:2021vvq}. These achievements further motivate people's interest in the studies of multi-quark states in the light hadron sector~\cite{Guo:2017jvc}. Apparently, the verification of the possible $\Sigma^*$ state is very meaningful and important for the examination of the pentaquark models and the clarification of the internal quark constituents of light hadrons.

The aforementioned $\Sigma^*(J^P=1/2^-)$ state is usually denoted as $\Sigma(1380)1/2^-$ and has been studied in many theoretical papers~\cite{Zou:2006uh,Zou:2007mk,Wu:2009tu,Wu:2009nw,Zou:2010zzb,Gao:2010hy,Chen:2013vxa,Xie:2014zga,Xie:2017xwx,Lyu:2024qgc,Huang:2024oai,Oller:2000fj,Oller:2006jw,Guo:2012vv,Khemchandani:2018amu,CLAS:2013rjt,Kim:2021wov,Khemchandani:2022dst}. The indication of its existence has been for the first time revealed in the $J/\psi$ decay process based on the lowest $J^P=1/2^-$ baryon nonet deduced from the diquark cluster picture~\cite{Zou:2006uh,Zou:2007mk}. Subsequently, Refs.~\cite{Wu:2009tu,Wu:2009nw,Zou:2010zzb} propose further evidence by fitting the experimental data of the $K^- p \to \Lambda \pi^+ \pi^-$ reaction, in the mean time determining its mass and width as $1380$ MeV and $120$ MeV, respectively. Furthermore, according to the results of Refs.~\cite{Gao:2010hy,Chen:2013vxa,Xie:2014zga}, the $\Sigma(1380)1/2^-$ plays an important role in reproducing the experimental data of the reactions $\gamma n \to K^+ \Sigma^-(1385)$ and $\Lambda p \to \Lambda p \pi^0$. References~\cite{Xie:2017xwx,Lyu:2024qgc,Huang:2024oai} also suggest to search for this state in the $\Lambda^+_c \to \eta \pi^+ \Lambda$ and $J/\psi \to \Lambda \bar{\Lambda} \pi^0$ decay processes by virtue of the appropriate observables. Recently, BESIII Collaboration reported the evidence for $\Sigma^+(1380)$ decaying into $\Lambda\pi^+$ with statistical significance larger than 3$\sigma$ in $\Lambda^+_c \to \eta \pi^+ \Lambda$ \cite{BESIII:2024mbf}. In addition to these studies, the unitary chiral perturbation theory predicted a similar isospin-one resonance with the mass around $1400$ MeV~\cite{Oller:2000fj,Oller:2006jw,Guo:2012vv,Khemchandani:2018amu}.\makeatletter\def\Hy@Warning#1{}\makeatother \footnote{According to the coupled-channel meson-baryon dynamics, there may exist two isospin-one poles in the energy region around $1400$ MeV, but it has not been experimentally confirmed.} After that, the nominal $\Sigma(1400)$ state has been thoroughly investigated in the photoproduction processes of light hyperons~\cite{CLAS:2013rjt,Kim:2021wov,Khemchandani:2022dst}. However, until now it remains unclear whether the $\Sigma(1380)$ and $\Sigma(1400)$ correspond to the same resonance and what their parameters are. Further investigations into the $\Sigma(1380)$ and/or $\Sigma(1400)$ states in various reactions with different approaches are necessary for understanding the $\Sigma^*$ spectroscopy.

Recently, triangle singularity (TS) proposed by Landau in 1959~\cite{Landau:1959fi} has once again attracted a lot of attention. Historically, Mandelstam has interpreted the anomalous behavior of the deuteron electromagnetic form factor by means of TS~\cite{Mandelstam:1960}. Nowadays, studies on this kinematic singularity offer numerous reasonable explanations for some important puzzles and the spectra~\cite{Wu:2011yx,Aceti:2012dj,Wu:2012pg,Achasov:2015uua,Du:2019idk,Jing:2019cbw,Guo:2019qcn,Sakai:2020ucu,Molina:2020kyu,Sakai:2020crh,Yan:2022eiy}. For instance, the reaction amplitude enhancement caused by the TS mechanism clarifies the abnormally large isospin-breaking effects observed in $J/\psi \to \gamma \eta(1405) \to \pi^0 f_0(980)$~\cite{Wu:2011yx,Aceti:2012dj,Wu:2012pg,Achasov:2015uua,Du:2019idk}. Furthermore, the TS mechanism also reproduces the band structure on the $\pi^0\phi$ distribution in the Dalitz plot of the decay $J/\psi \to \eta\pi^0\phi$~\cite{Jing:2019cbw}. Based on the fact that the occurrence of TS is highly sensitive to the kinematic conditions, the mass of $X(3872)$ can be precisely determined~\cite{Guo:2019qcn,Sakai:2020ucu,Molina:2020kyu,Sakai:2020crh,Yan:2022eiy}. On the other hand, our recent studies have found that the spin effects provide an alternative and useful tool for verifying the TS mechanism~\cite{Wang:2022wdm,Wang:2023xua}. In a word, the TS mechanism plays an essential role in searching for the possible resonances.

\begin{figure}[tbp]
    \begin{center}
        \includegraphics[scale=0.5]{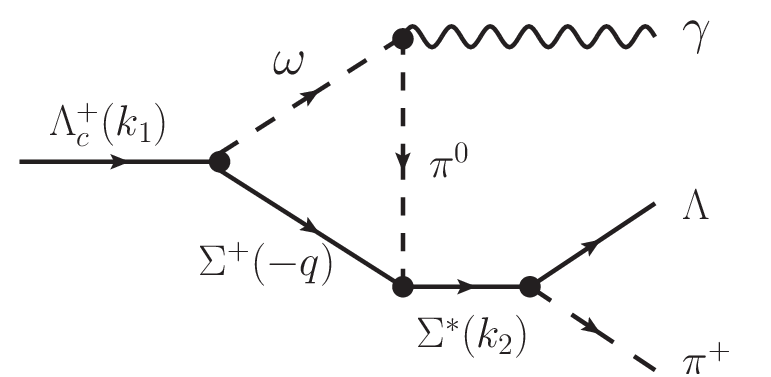}
        \caption{The Feynman diagram for the radiative decay process $\Lambda^+_c \to \gamma \pi^+ \Lambda$ through a triangle loop involving the $\omega$, $\Sigma$, and $\pi$. Here $\Sigma^*$ denotes the $\Sigma(1380)1/2^-$ or $\Sigma(1385)3/2^+$.} \label{FeynmanTS}
    \end{center}
\end{figure}

The $\Sigma(1380)1/2^-$ has a similar mass to the $\Sigma(1385)3/2^+$ state, although it has a larger width than the latter, which adds the difficulty to distinguish the signal of $\Sigma(1380)1/2^-$ from the experimental data in a traditional way. In this work, we propose the study of the $\Sigma(1380)1/2^-$ state through the TS mechanism in the radiative decay process $\Lambda^+_c \to \gamma \pi^+ \Lambda$, which occurs via the triangle loop involving $\omega$, $\Sigma^+$, and $\pi^0$ as illustrated in Fig.~\ref{FeynmanTS}. The particles $\Sigma^+$ and $\pi^0$ couple to the intermediate state $\Sigma^*$ \footnote{Here we only consider the contributions from two $\Sigma^*$ states in the studied energy region: $\Sigma(1385)3/2^+$ and $\Sigma(1380)1/2^-$.}, which subsequently decays into the final particles $\pi^+$ and $\Lambda$. The peak caused by the TS appears in the $\pi\Lambda$ invariant mass spectrum. Furthermore, according to Coleman-Norton theorem~\cite{Coleman:1965xm}, a TS occurs only when the subprocesses of the triangle loop take place in a classical manner: all internal particles are on shell simultaneously with their three momenta being collinear in the rest frame of the mother particle. In Fig.~\ref{FeynmanTS}, the internal particles are near their thresholds at the TS position. This means that when the TS occurs, the values of the three momenta of internal particles are rather small, which consequently suppresses the amplitudes with non-$S$-wave coupling vertices compared to the $S$-wave vertex~\cite{Guo:2019twa}. Because the $\Sigma(1380)1/2^-$ and $\Sigma(1385)3/2^+$ couple to $\pi\Lambda$ in $S$ wave and $P$ wave, respectively, the presence of $\Sigma(1380)1/2^-$ in $\Lambda^+_c \to \gamma \pi^+ \Lambda$ does amplify the strength of TS in this decay process~\cite{Huang:2024oai}. In other words, instead of a non-prominent Breit-Wigner peak, a significant sharp peak from the TS mechanism indicates the existence of $\Sigma(1380)1/2^-$, though other possible tree-level processes may also have influences on the line shape. 

This paper is organized as follows. In Sec.~\ref{model}, we present the theoretical framework and amplitudes for the reaction $\Lambda^+_c \to \gamma \pi^+ \Lambda$. In Sec.~\ref{results}, we show the numerical results and discuss their implications. Finally, we summarize our findings and conclusions in Sec.~\ref{summary}.

\section{MODEL AND INGREDIENTS}
\label{model}

In this work, we introduce the TS mechanism to study the possible $\Sigma(1380)1/2^-$ state in the radiative decay process $\Lambda^+_c \to \gamma \pi^+ \Lambda$ within an effective Lagrangian approach. According to the results in Ref.~\cite{Bayar:2016ftu}, the production of a TS must satisfy some special kinematic conditions. If the nominal masses in PDG~\cite{ParticleDataGroup:2022pth} are adopted for the involved particles in Fig.~\ref{FeynmanTS}, the TS occurs at $M_{\pi^+\Lambda}=1.348$ GeV.

The relevant Lagrangian densities should be introduced in order to calculate the amplitudes of Fig.~\ref{FeynmanTS}. For the weak decay vertex $\Lambda^+_c \to \omega\Sigma^+$, we decompose the decay amplitude into two different structures due to the parity violation in the weak interaction~\cite{Xie:2017xwx,Xie:2017erh,Liu:2019dqc}
\begin{equation}
    \CM(\Lambda^+_c \to \Sigma^+ \omega) = i \bar{u}_{\sm{\Sigma}} \gamma^\mu \varepsilon^*_\mu \left( g_{\sm{A}} + g_{\sm{B}} \gamma_5 \right) u_{\sm{\Lambda_c}},
\end{equation}
where $g_{\sm{A}}$ and $g_{\sm{B}}$ denote the coupling constants for different amplitude structures. Generally, their values can be calculated within some theoretical models~\cite{Cheng:1991sn,Lu:2009cm} or determined by fitting the experimental data. In our calculations, the shape of the invariant mass spectrum of final $\pi\Lambda$ is insensitive to these couplings. For simplicity, we take $g_{\sm{A}}=g_{\sm{B}}=g_{\sm{\Lambda_c \Sigma \omega}}=4.70\times10^{-7}$ in this work, which is pinned down through the relevant partial decay width in PDG. For the other electromagnetic and strong decay vertices, we adopt the effective Lagrangians as follows~\cite{Xie:2014zga,Lu:2014yba,Liu:2017acq,Zhao:2019syt,Fan:2019lwc,Xie:2014kja} (omitting the isospin structures):
\begin{eqnarray}
    \CL_{\omega \gamma \pi} &=& e\frac{g_{\omega \gamma \pi}}{m_\omega} \varepsilon^{\mu \nu \alpha \beta} \left(\partial_\mu \omega_\nu\right) \left(\partial_\alpha A_\beta\right)  \pi, \\
    \CL_{\Sigma^{*}_1 \pi Y} &=& g_{\sm{\Sigma^*_1 \pi Y}} \bar{Y} \pi \Sigma^{*}_1 + \text{H.c.}, \\
    \CL_{\Sigma^{*}_2 \pi Y} &=& \frac{g_{\sm{\Sigma^*_2 \pi Y}}}{m_\pi} \bar{Y} \left(\partial_\mu \pi\right)\Sigma^{*\mu}_2 + \text{H.c.},
\end{eqnarray}
where $A$, $\Sigma^*_1$, $\Sigma^*_2$, and $Y$ represent the photon field, $\Sigma(1380)1/2^-$, $\Sigma(1385)3/2^+$, and hyperon ($\Lambda$ or $\Sigma$), respectively. The value of $e$ is taken as $\sqrt{4\pi/137}$. The coupling constants $g_{\omega \gamma \pi}$ and $g_{\Sigma^* \pi Y}$ appearing in the above Lagrangian densities can be determined through the corresponding partial decay widths using
\begin{eqnarray}
    \Gamma_{\omega \to \pi \gamma} &=& e^2\frac{g^2_{\omega \gamma \pi}} {12\pi} \frac{|\boldm{p}_{\pi}|^3}{m^2_{\omega}}, \\
    \Gamma_{\Sigma^*_1 \pi Y} &=& \kappa\frac{ g_{\sm{\Sigma^*_1 \pi Y}}^{2}} {4 \pi} \frac{E_{\sm{Y}}+m_{\sm{Y}}}{m_{\sm{\Sigma^*_1}}} |\boldm{p}_{\pi}|, \\
    \Gamma_{\Sigma^*_2 \pi Y} &=& \kappa\frac{ g_{\sm{\Sigma^*_2 \pi Y}}^{2}} {12 \pi} \frac{E_{\sm{Y}}+m_{\sm{Y}}}{m_{\sm{\Sigma^*_2}} m_{\pi}^{2}} |\boldm{p}_{\pi}|^3,
\end{eqnarray}
where $\boldm{p}_{\pi}$ denotes the three momentum of the pion; $E_{\sm{Y}}$ represents the hyperon energy in the rest frame of $\Lambda^+_c$; and $\kappa$ is a constant which equals $1$ for $Y=\Lambda$ and $2$ for $Y=\Sigma$~\cite{Xie:2014kja}. We follow Refs.~\cite{Wu:2009tu,Wu:2009nw,Zou:2010zzb} and assume the mass and total width of $\Sigma(1380)1/2^-$ as $1380$ MeV and $120$ MeV, respectively. However, there are no literature results on its branching ratios to different channels. Here we tentatively take branching ratios of $\Sigma(1380)1/2^-$ the same as those of $\Sigma(1385)3/2^+$, which lead to the couplings $g_{\sm{\Sigma^*_1 \pi Y}}$ summarized in Table~\ref{tab1}. The dependency of the results on the branching ratios will be discussed in the next section.

According to the above Lagrangian densities, the invariant decay amplitudes for the triangle loop diagram in Fig.\ref{FeynmanTS} are: 
\begin{widetext}
\begin{eqnarray}
    \CM^{TS}_1 &=& g^{\sm{TS}}_1 \bar{u}_{\sm{\Lambda}} G^{\frac{1}{2}}(p_{\sm{\Sigma^*_1}}) F(p_{\sm{\Sigma^*_1}}) \int\frac{\text{d}^4q}{(2\pi)^4} G^{\frac{1}{2}}(p_{\sm{\Sigma}}) \gamma_\sigma \epsilon_{\mu\nu\alpha\beta} p^\mu_\omega G^{1,\nu\sigma}(p_\omega) p^\alpha_\gamma \varepsilon^{*\beta}_{\gamma} G^0(p_{\pi^0}) F(p_{\pi^0}) \left( 1+\gamma_5 \right) u_{\sm{\Lambda_c}} \nonumber \\
    &\equiv& g^{\sm{TS}}_1 \bar{u}_{\sm{\Lambda}} G^{\frac{1}{2}}(p_{\sm{\Sigma^*_1}}) \CM_{1,\mu} \varepsilon^{*\mu}_{\gamma} \left( 1+\gamma_5 \right) u_{\sm{\Lambda_c}} F(p_{\sm{\Sigma^*_1}}), \label{AmpTS1} \\
    \CM^{TS}_2 &=& g^{\sm{TS}}_2 \bar{u}_{\sm{\Lambda}} p^\lambda_{\pi^+} G^{\frac{3}{2}}_{\lambda\rho}(p_{\sm{\Sigma^*_2}}) F(p_{\sm{\Sigma^*_2}}) \int\frac{\text{d}^4q}{(2\pi)^4} p^\rho_{\pi^0} G^{\frac{1}{2}}(p_{\sm{\Sigma}}) \gamma_\sigma \epsilon_{\mu\nu\alpha\beta} p^\mu_\omega G^{1,\nu\sigma}(p_\omega) p^\alpha_\gamma \varepsilon^{*\beta}_{\gamma} G^0(p_{\pi^0}) F(p_{\pi^0}) \left( 1+\gamma_5 \right) u_{\sm{\Lambda_c}} \nonumber \\
    &\equiv& g^{\sm{TS}}_2 \bar{u}_{\sm{\Lambda}} p^\lambda_{\pi^+} G^{\frac{3}{2}}_{\lambda\rho}(p_{\sm{\Sigma^*_2}}) \CM^\rho_{2,\mu} \varepsilon^{*\mu}_{\gamma} \left( 1+\gamma_5 \right) u_{\sm{\Lambda_c}} F(p_{\sm{\Sigma^*_2}}), \label{AmpTS2}
\end{eqnarray}
\end{widetext}
where $u_\Lambda$ and $u_{\Lambda_c}$ are the spinors of the $\Lambda$ and $\Lambda_c$, respectively; $\varepsilon^*_\gamma$ is the spin polarization vector of the photon; and $g^{\sm{TS}}_1$ and $g^{\sm{TS}}_2$ read
\begin{eqnarray}
g^{\sm{TS}}_1 &=& e \frac{i}{m_\omega} g_{\sm{\Lambda_c \Sigma \omega}} \, g_{\omega \gamma \pi} \, g_{\sm{\Sigma^*_1 \pi \Sigma}} \, g_{\sm{\Sigma^*_1 \pi \Lambda}}, \\
g^{\sm{TS}}_2 &=& e \frac{i}{m_\omega m^2_\pi} g_{\sm{\Lambda_c \Sigma \omega}} \, g_{\omega \gamma \pi} \, g_{\sm{\Sigma^*_2 \pi \Sigma}} \, g_{\sm{\Sigma^*_2 \pi \Lambda}}.
\end{eqnarray}
The propagators of the intermediate particles with spin $J$ are denoted by the symbol $G^J$~\cite{Fan:2019lwc,Chen:2020szc,Xie:2013db}
\begin{eqnarray}
    G^0(p) &=& \frac{i}{p^2-m^2}\label{pp1}, \\
    G^1_{\mu\nu}(p) &=& -\frac{i \left(g_{\mu\nu}-p_\mu p_\nu/m^2\right)}{p^2-m^2+im\Gamma}\label{pp2}, \\
    G^{\frac{1}{2}}(p) &=& \frac{i \left(\slashed{p}+m\right)}{p^2-m^2}\label{pp3}, \\
    G^{\frac{3}{2}}_{\mu\nu}(p) &=& \frac{i \left(\slashed{p}+m\right)}{p^2-m^2+im\Gamma}\left[ -g_{\mu\nu} + \frac{1}{3}\gamma_\mu \gamma_\nu \right. \nonumber \\ & & \left. + \, \frac{1}{3m}\left(\gamma_\mu p_\nu - \gamma_\nu p_\mu \right) +\frac{2}{3m^2}p_\mu p_\nu \right]\label{pp4},
\end{eqnarray}
where $p$ and $m$ are the four momentum and mass of the intermediate state, respectively, and $\Gamma$ is the corresponding constant width. The masses and widths of the involved hadrons other than $\Sigma(1380)1/2^-$ are taken from PDG~\cite{ParticleDataGroup:2022pth}. Meanwhile, to avoid the loop integral divergence and consider the off-shell effects of the intermediate hadrons simultaneously, we introduce the phenomenological form factor $F(p)$ for the intermediate $\pi$ meson and the $\Sigma^*$ resonances~\cite{Skoupil:2016ast,Fatima:2020tyh,Huang:2020kxf}
\begin{equation}
    F(p) = \frac{\Lambda^4}{\Lambda^4+(p^2-m^2)^2},
\end{equation}
where $\Lambda$ is the cutoff parameter. Since the off-shell effects of the intermediate $\omega$ and $\Sigma^+$ are small near the TS, their form factors are not included here. In our calculations, we adopt the cutoff values given by the empirical formula $\Lambda=m+\alpha \Lambda_{\text{QCD}}$~\cite{Xiao:2018kfx,Huang:2020kxf,Ling:2021lmq}, where $\Lambda_{\text{QCD}} = 0.22$ GeV and $\alpha$ is set to unity.

\begin{figure}[tbp]
    \begin{center}
        \includegraphics[scale=0.5]{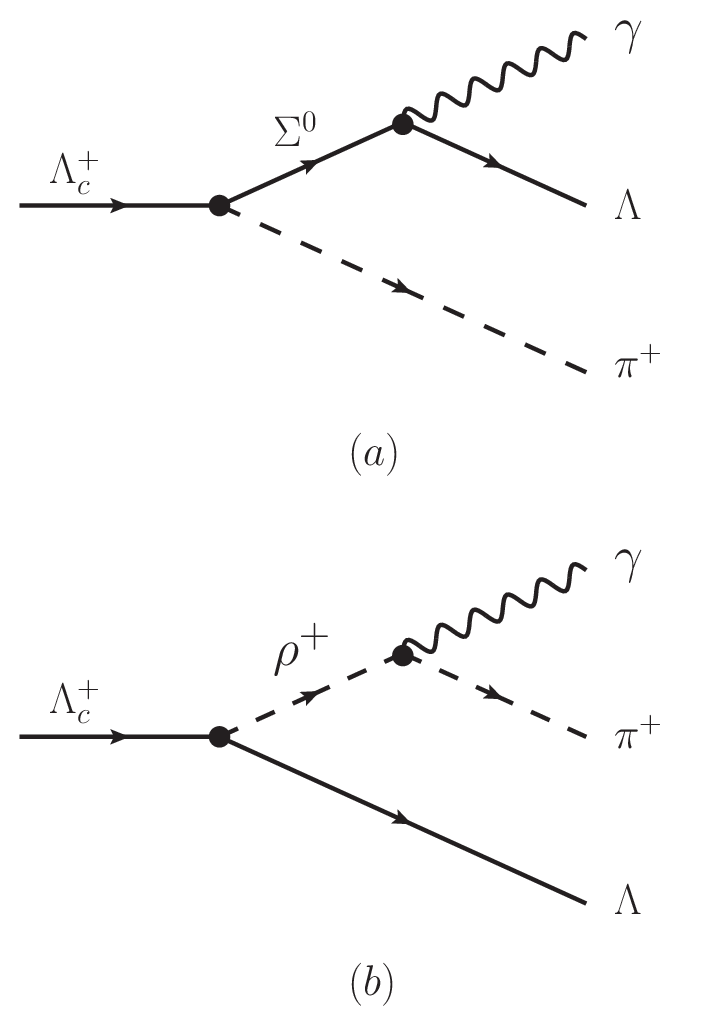}
        \caption{The possible tree-level diagrams for the radiative decay process $\Lambda^+_c \to \gamma \pi^+ \Lambda$.}
        \label{FeynmanBack}
    \end{center}
\end{figure}

As mentioned in the Introduction, the contributions from tree-level processes may play an important role in the radiative decay process $\Lambda^+_c \to \gamma \pi^+ \Lambda$ and even change the shape of the TS structure in the $\pi\Lambda$ invariant mass distributions. Based on the available experimental information, the tree-level diagrams plotted in Fig.~\ref{FeynmanBack} may contribute significantly to the decay amplitudes. The weak decay vertices therein can also be expressed as follows
\begin{eqnarray}
    \CM(\Lambda^+_c \to \Sigma^0 \pi^+) &=& i g_{\sm{\Lambda_c \Sigma \pi}} \bar{u}_{\sm{\Sigma}} \left( 1 + \gamma_5 \right) u_{\sm{\Lambda_c}}, \\
    \CM(\Lambda^+_c \to \Lambda \rho^+) &=& i g_{\sm{\Lambda_c \Lambda \rho}} \bar{u}_{\sm{\Lambda}} \gamma^\mu \varepsilon^*_\mu \left( 1 + \gamma_5 \right) u_{\sm{\Lambda_c}}.
\end{eqnarray}

For the radiative decay vertices, their effective Lagrangian densities are taken from Refs.~\cite{Fan:2019lwc,Wang:2017tpe}
\begin{eqnarray}
    \CL_{\Sigma \Lambda \gamma} &=& e \frac{\kappa_{\Sigma \Lambda}}{2 m_{N}} \bar{\Lambda} \sigma^{\mu \nu} \left( \partial_{\nu} A_{\mu} \right) \Sigma + \mathrm{H.c},\\
    \CL_{\rho \gamma \pi} &=& e \frac{g_{\rho \gamma \pi}}{m_\rho} \varepsilon^{\mu \nu \alpha \beta} \left(\partial_\mu \rho_\nu\right) \left(\partial_\alpha A_\beta\right) \pi,
\end{eqnarray}
where $\kappa_{\Sigma \Lambda}=-1.61$ represents the anomalous magnetic moment, and the coupling constant $g_{\rho \gamma \pi}=0.56$ is determined by the partial decay width of $\rho\to \gamma\pi$. Finally, we obtain two tree-level invariant decay amplitudes
\begin{eqnarray}
    \CM^{Tr}_a &=& g^{\sm{Tr}}_a \bar{u}_{\sm{\Lambda}} \sigma_{\mu \nu} p^\mu_\gamma \varepsilon^{*\nu}_{\gamma} G^{\frac{1}{2}}(p_{\sm{\Sigma}}) \left( 1+\gamma_5 \right) u_{\sm{\Lambda_c}} \nonumber \\ 
    & & \times \,  F(p_{\sm{\Sigma}}), \label{AmpTree1} \\
    \CM^{Tr}_b &=& g^{\sm{Tr}}_b \bar{u}_{\sm{\Lambda}} \gamma_\lambda \epsilon_{\mu\nu\alpha\beta} p^\mu_\rho G^{1,\nu\lambda}(p_\rho) p^\alpha_\gamma \varepsilon^{*\beta}_{\gamma} \nonumber \\ 
    & & \times \, \left( 1+\gamma_5 \right) u_{\sm{\Lambda_c}} F(p_{\rho}), \label{AmpTree2}
\end{eqnarray}
where
\begin{eqnarray}
g^{\sm{Tr}}_a &=& -e \frac{\kappa_{\Sigma \Lambda}}{2 m_{N}} g_{\sm{\Lambda_c \Sigma \pi}},  \\
g^{\sm{Tr}}_b &=& e \frac{i}{m_\rho} g_{\sm{\Lambda_c \Lambda \rho}} \, g_{\rho \gamma \pi}.
\end{eqnarray}

To investigate the effects of $\Sigma(1380)1/2^-$ on the TS, we construct and compare the following two parameterizations of the total amplitudes: 
\begin{eqnarray}
    \CM^{Total}_{I} &=& \CM^{TS}_2 + \left( \CM^{Tr}_a + \CM^{Tr}_b \right) e^{i\phi}, \label{Total1} \\
    \CM^{Total}_{II} &=& \CM^{TS}_1 + \CM^{TS}_2 + \left( \CM^{Tr}_a + \CM^{Tr}_b \right) e^{i\phi}, \label{Total2}
\end{eqnarray}
where $\CM^{Total}_{I}$ and $\CM^{Total}_{II}$ correspond to the scenarios without and with the contribution of $\Sigma(1380)1/2^-$, respectively. To analyze the interference effects, we introduce a phase $\phi$ between the TS and tree-level amplitudes. The $\pi\Lambda$ invariant mass distribution for both scenarios can be calculated with the following formula~\cite{ParticleDataGroup:2022pth,Jing:2019cbw}
\begin{equation}\label{Distribution}
    \frac{\text{d}\Gamma}{\text{d} M_{\pi\Lambda}} \! = \! \frac{4m_\Lambda}{(2 \pi)^{5} 2^4 m_{\sm{\Lambda_c}}} \frac{\left|\boldm{p}_{\gamma}\right| \! \left|\boldm{p}_{\pi}^{*}\right|}{2} \! \int \! \text{d} \Omega_{\gamma} \text{d} \Omega_{\pi}^{*} \sum_{\text{spin}} \left|\CM\right|^{2} \!,
\end{equation}
where the quantities with superscript ``$*$'' are in the center of mass frame of $\pi^+\Lambda$, while the others are in the $\Lambda^+_c$ rest frame.

\begin{table}[tbp]
    \caption{Coupling constants used in this work. The decay widths and the branching ratios of the states other than $\Sigma(1380)1/2^-$ are taken from Ref.~\cite{ParticleDataGroup:2022pth}. The mass and  width of $\Sigma(1380)1/2^-$ are set as $1380$ MeV and $120$ MeV, respectively, according to Refs.~\cite{Wu:2009tu,Wu:2009nw,Zou:2010zzb}. Moreover, we assume the branching ratios of $\Sigma(1380)1/2^-$ are the same as those of $\Sigma(1385)3/2^+$. }
    \begin{tabular*}{\columnwidth}{@{\extracolsep\fill}ccccc}
        \hline\hline
        State         & Width                &  Decay           &      Branching      &    $g$  \\[-3pt]
                        & (GeV)                &  channel         &        ratio            &          \\
        \hline
        $\Lambda^+_c$ & $3.29\times10^{-12}$ & $\Sigma^+\omega$ & $1.70\times10^{-2}$ & $4.70\times10^{-7}$   \\
                      &                      & $\Sigma^0\pi^+$  & $1.27\times10^{-2}$ & $7.08\times10^{-7}$   \\
                      &                      & $\Lambda\rho^+$ & $4.00\times10^{-2}$ & $6.55\times10^{-7}$   \\
        $\omega$      & $8.68\times10^{-3}$  & $\pi^0\gamma$    & $8.35\times10^{-2}$ & 1.83   \\
        $\rho$        & 0.149                & $\pi^+\gamma$    & $4.50\times10^{-4}$ & 0.56   \\
        $\Sigma(1380)$&  0.12                & $\Sigma\pi$      & 0.117               & 0.64   \\
                      &                      & $\Lambda\pi$     & 0.870               & 1.98   \\
        $\Sigma(1385)$& $3.62\times10^{-2}$  & $\Sigma\pi$      & 0.117               & 0.63   \\
                      &                      & $\Lambda\pi$     & 0.870               & 1.22   \\
        \hline\hline
    \end{tabular*}
    \label{tab1}
\end{table}

\section{RESULTS AND DISCUSSION}
\label{results}

For the radiative decay reaction $\Lambda^+_c \to \gamma \pi^+ \Lambda$, the loop decay process shown in Fig.~\ref{FeynmanTS} leads to a TS peak on the $\pi\Lambda$ invariant mass spectrum. In this section, we shall explore the possibility of detecting the $\Sigma(1380)1/2^-$ resonance through the TS. The loop integrals in Eqs.~\eqref{AmpTS1} and \eqref{AmpTS2} can be numerically calculated by using the package LoopTools~\cite{Hahn:2000jm}.

\begin{figure}[tbp]
    \begin{center}
        \includegraphics[scale=0.31]{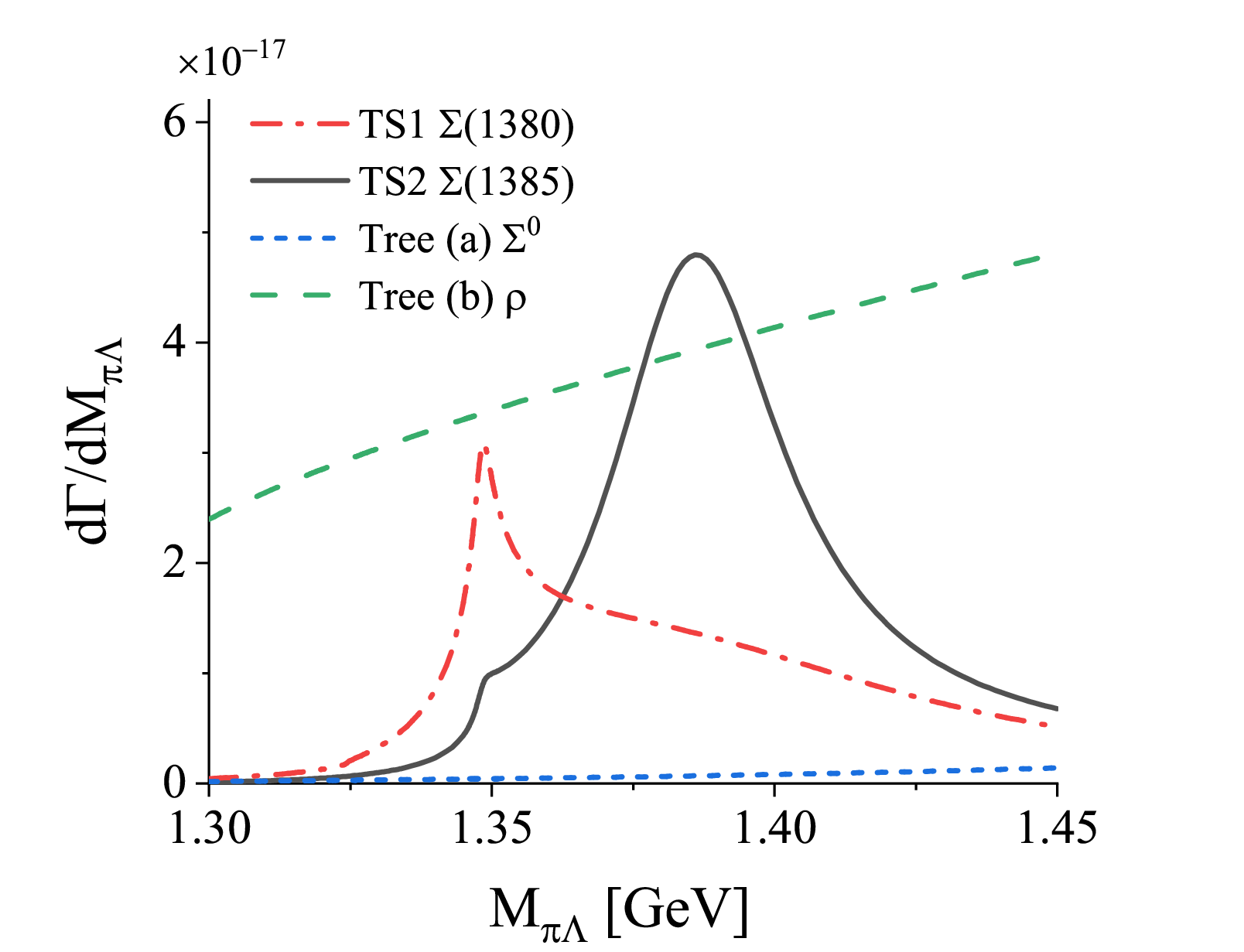}
        \caption{The distribution of the differential decay width versus the invariant mass $M_{\pi \Lambda}$ in $\Lambda_c \to \gamma \pi \Lambda$. The red dot-dashed, black solid, blue dotted, and green dashed lines denote the results of $\Sigma(1380)1/2^-$, $\Sigma(1385)3/2^+$, $\Sigma^0$, and $\rho$ as the intermediate states, respectively.}
        \label{MS1}
    \end{center}
\end{figure}

\begin{figure*}[tbp]
    \begin{center}
        \includegraphics[scale=0.6]{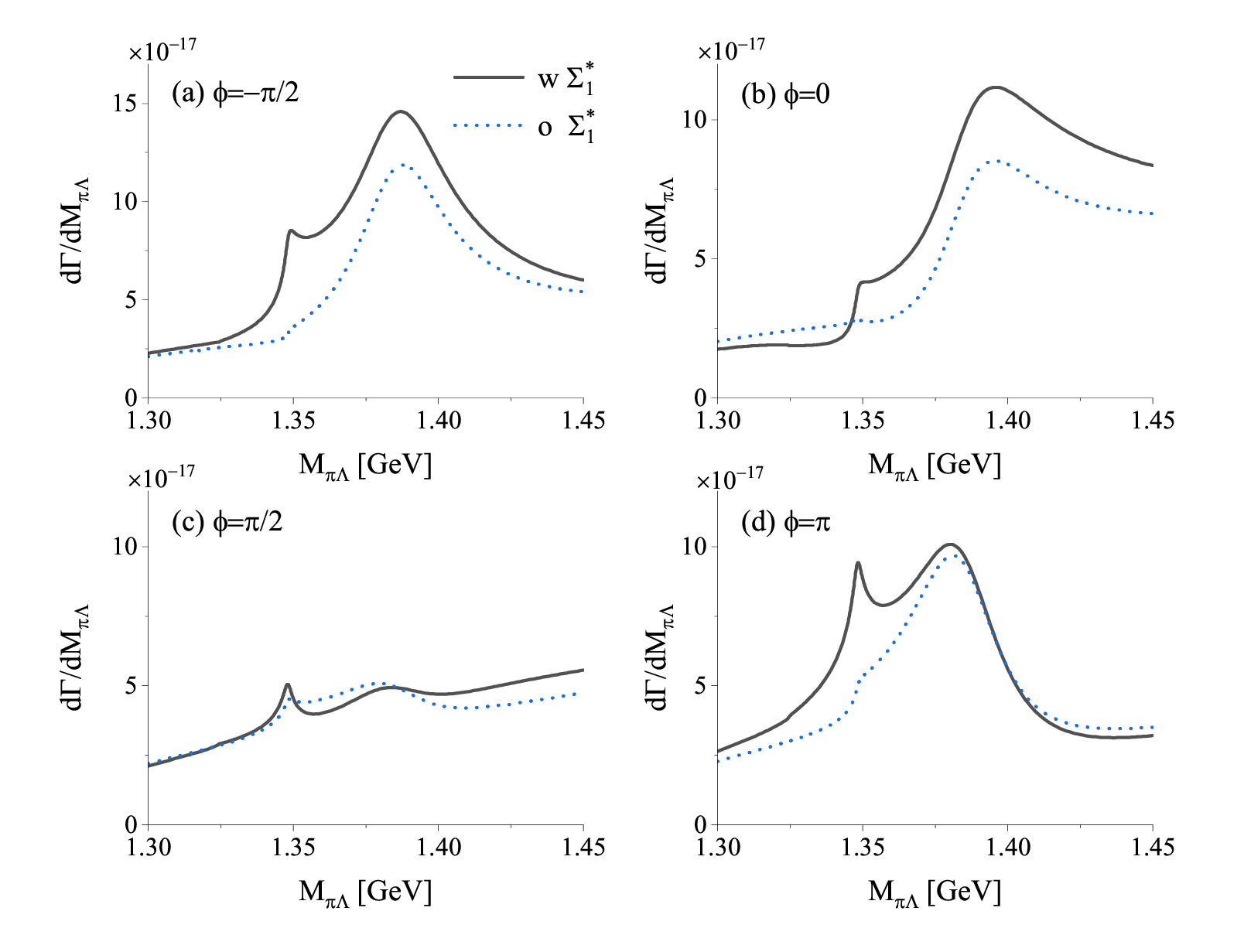}
        \caption{The distribution of the differential decay width versus the invariant mass $M_{\pi \Lambda}$ for the case with/without (w/o) the contribution of $\Sigma(1380)1/2^-$. The different sub-figures denote the results with taking the phase angle $\phi = -\pi/2, 0, \pi/2,$ and $\pi$, respectively.}
        \label{MS2}
    \end{center}
\end{figure*}

By use of Eq.~\eqref{Distribution}, we obtain the distributions of the differential decay width versus the invariant mass $M_{\pi\Lambda}$. For the individual amplitudes shown in Eqs.~\eqref{AmpTS1}, \eqref{AmpTS2}, \eqref{AmpTree1}, and \eqref{AmpTree2}, the obtained numerical results are depicted in Fig.~\ref{MS1}, which emphasize their own features that might be smeared or weakened in total amplitude with interference effects. As can be seen from the figure, $\Sigma(1385)3/2^+$ is recognized as a noticeable Breit-Wigner peak, whereas the structure of $\Sigma(1380)1/2^-$ is unobvious due to the big width. Although both the loop processes including $\Sigma(1380)1/2^-$ and $\Sigma(1385)3/2^+$ can generate the TS signal around $M_{\pi\Lambda}=1.35$ GeV, their structures exhibit distinguishing features: the TS structure with $\Sigma(1380)1/2^-$ appears as a sharp peak, while that with $\Sigma(1385)3/2^+$ only shows a small shoulder. As already mentioned, this discrepancy stems from the $P$-wave vertices of $\Sigma(1385)3/2^+$, which are proportional to the small momentum at the TS~\cite{Huang:2024oai,Guo:2019twa}. Although alternative processes such as higher excited resonances may also provide $S$-wave couplings, the mass of the possible $\Sigma(1380)1/2^-$ is the closest to the TS. Therefore, we expect that a sharp TS peak could indicate the presence of the $\Sigma(1380)1/2^-$ state. On the other hand, since the off-shell effect of the intermediate state $\Sigma^0$ plotted in Fig.~\ref{FeynmanBack} significantly suppresses the tree-level contribution within the studied energy region, here we can ignore this process by making a suitable kinematic cut on the invariant mass $M_{\pi\Lambda}$. On the contrary, the tree-level process involving the intermediate $\rho$ meson must be considered, as its contribution exhibits comparable strength to the triangle loop, and moreover, the broad width of the $\rho$ meson makes this contribution hard to be separated in the Dalitz plot.

Coming to the total amplitudes, Fig.~\ref{MS2} shows the mass distributions of the final state $\pi\Lambda$ for the two scenarios with or without $\Sigma(1380)1/2^-$, under different values of the phase $\phi$. As expected, the interference effects between the TS and tree-level amplitudes indeed visibly change the shape of the TS peak and $\Sigma(1385)3/2^+$ structure, compared to the individual plots in Fig.~\ref{MS1}. Nevertheless, the Breit-Wigner peak of $\Sigma(1380)1/2^-$ is still not manifest. Instead, just like the individual case, the $\Sigma(1380)1/2^-$ is featured by the pronounced sharp TS peak at $M_{\pi\Lambda}\approx 1.35$ GeV in contrast to the case in which it is absent. This can be seen in Fig.~\ref{MS2} as the phase changes. For a large range of the $\phi$ value, that TS peak is obvious in the curve, together with an enhancement of the magnitude of the differential decay width. Only when $\phi$ is close to $\pi/2$, the distinction is weakened, though the TS peak with $\Sigma(1380)1/2^-$ is still more recognizable. In fact, since the TS peak is known to be located around $1.35$ GeV, high-precision experimental measurements around that point would surely help figure out whether $\Sigma(1380)1/2^-$ exists in a quantitative manner. For example, a comparison between the fit qualities to the data of the two scenarios provides the statistical significance of $\Sigma(1380)1/2^-$. 

Constrained by the particle masses, $\Sigma(1380)1/2^-$ only has two possible strong decay channels, $\pi \Sigma$ and $\pi \Lambda$, which nearly contribute to $100\%$ of the decay width. All conclusions drawn above concerning the TS with $\Sigma(1380)1/2^-$ are based on the assumption that the $\Sigma(1380)1/2^-$ has the same branching ratios to these two channels as the $\Sigma(1385)3/2^+$. However, this assumption has not been confirmed by experiments. Note that the strength of the TS with intermediate $\Sigma(1380)1/2^-$ is proportional to the product of those two branching ratios as depicted in Fig.~\ref{FeynmanTS}. Obviously, when both the branching ratios equal $50\%$, that process is maximally enhanced. In this work, even if we choose the branching ratios as ${\rm Br}(\pi \Lambda)\simeq 90\%$ and ${\rm Br}(\pi \Sigma)\simeq 10\%$ (see Table~\ref{tab1}), deviating much from $50\%$, the TS peak is still prominent. In fact this choice may likely underestimate the significance of the TS peak, unless $\Sigma(1380)1/2^-$ completely decays to one channel. 

\begin{figure}[tbp]
    \begin{center}
        \includegraphics[scale=0.5]{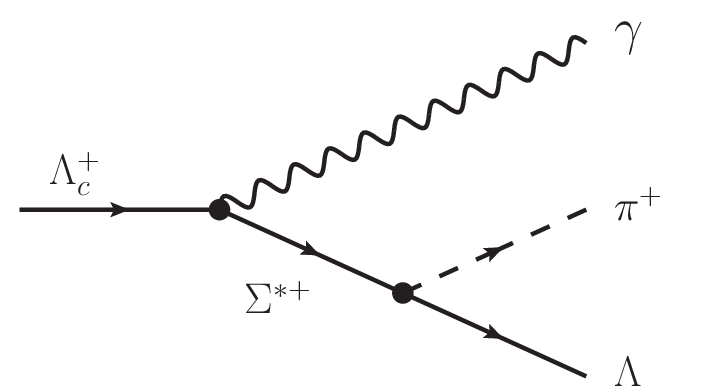}
        \caption{The radiative decay process $\Lambda^+_c \to \gamma \pi^+ \Lambda$ where the intermediate $\Sigma^*$ resonances are produced through the tree-level diagrams.}
        \label{FeynmanTree}
    \end{center}
\end{figure}

\begin{figure}[htbp]
    \begin{center}
        \includegraphics[scale=0.31]{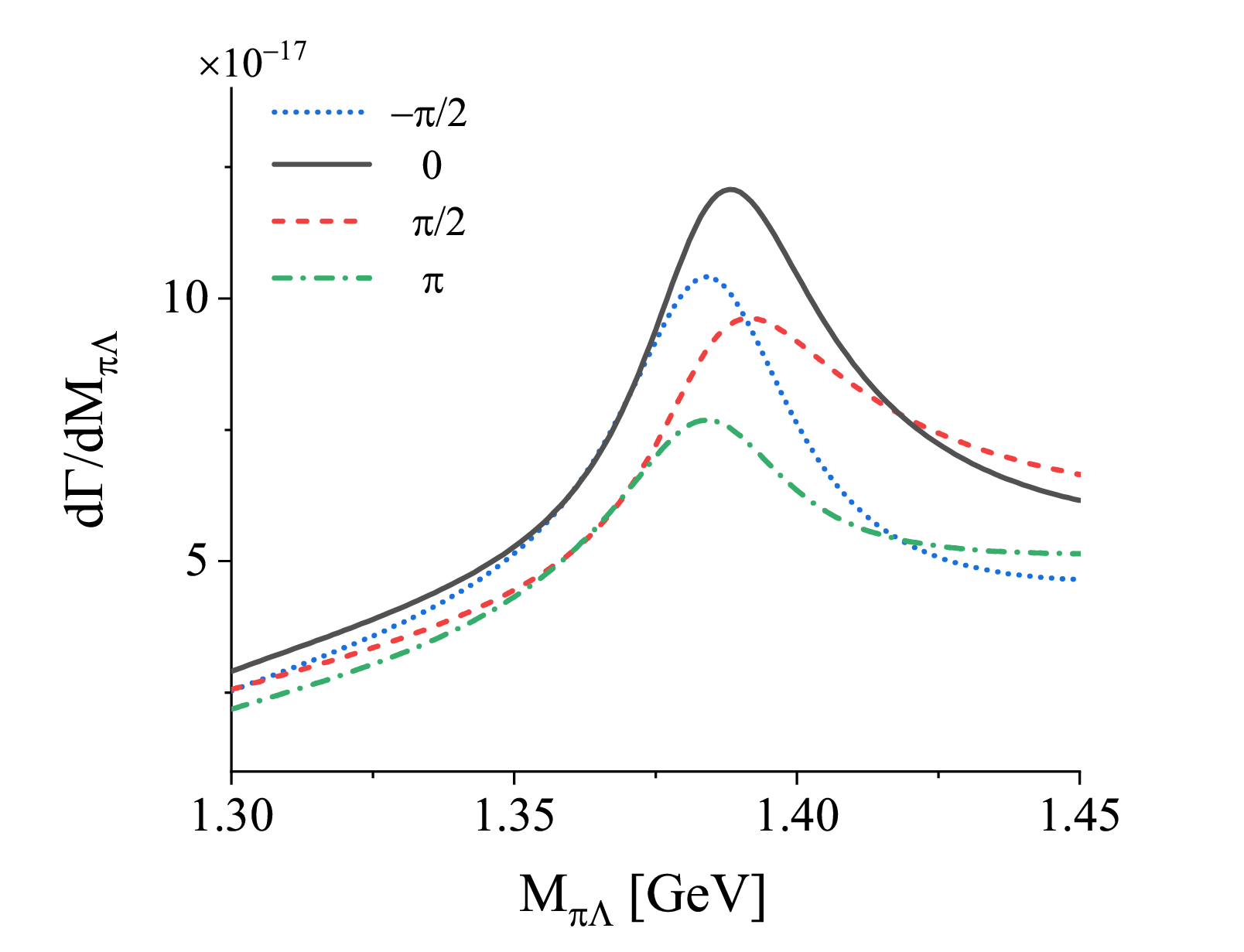}
        \caption{The distribution of the differential decay width versus the invariant mass $M_{\pi \Lambda}$ for the scenario that the contributions of the intermediate $\Sigma(1380)1/2^-$ and $\Sigma(1385)3/2^+$ states are generated by the tree-level diagram processes.}
        \label{MS3}
    \end{center}
\end{figure}

In addition, another source of the uncertainties can be the possible tree-level diagram corresponding to the radiative decay of the $\Lambda^+_c$ directly to the photon and the $\Sigma^*$, as shown in Fig.~\ref{FeynmanTree}. Unfortunately, the information about this weak vertex is totally unknown: there is neither theoretical predictions on the $\Lambda^+_c\Sigma(1380)\gamma$ coupling, nor experimental observations of the $\Lambda^+_c\to\Sigma(1385)\gamma$ process. Note that the latter may indicate a very small $\Lambda^+_c\to\Sigma(1385)\gamma$ coupling. Here we tentatively assign the values of the couplings, such that the magnitude of Fig.~\ref{FeynmanTree} is comparable with the triangle diagram Fig.~\ref{FeynmanTS}, see Fig.~\ref{MS3}. For every value of the phase, such kind of contributions only contain the Breit-Wigner peak of $\Sigma(1385)1/2^-$. Around the point $M_{\pi\Lambda}=1.35$ GeV, the behavior of this tree diagram is rather mild and can be regarded as a part of the background. We do not believe this effect can totally overthrow our main conclusion, unless the coupling of $\Lambda^+_c\Sigma(1380)\gamma$ is exaggeratedly large. 

We can also calculate the decay branching ratio of the reaction $\Lambda^+_c \to \gamma \pi^+ \Lambda$ using Eq.~\eqref{Distribution} with different phase angle values. The results are listed in the Table~\ref{tab2}, omitting the contribution from Fig.~\ref{FeynmanBack} (a) for the reason already discussed. We find the branching ratio for the process involving the loop diagram has a magnitude about $1.5\times10^{-5}$. This value, as a consequent of the $\Sigma^*$ resonance poles as well as the TS, is even comparable with the branching ratios of some strong decays in PDG. Therefore, it might be accessible to the experimental measurements.

\begin{table}[tbp]
    \caption{The decay branching ratio for $\Lambda^+_c \to \gamma \pi^+ \Lambda$ process when the phase angle $\phi$ is taken different values.}
    \begin{tabular*}{\columnwidth}{@{\extracolsep\fill}ccccc}    
        \hline\hline
        $\phi$ & $-\pi/2$  &  0 & $\pi/2$ & $\pi$ \\
        \hline
        Br($\times10^{-5}$)  & 1.53 & 1.90 & 1.60  & 1.22   \\
        \hline\hline
    \end{tabular*}
    \label{tab2}
\end{table}

\section{SUMMARY}
\label{summary}
In this paper, we propose a criterion for the existence of the possible $\Sigma(1380)1/2^-$ state, based on the triangle loop diagram of the $\Lambda^+_c \to \gamma \pi^+ \Lambda$ reaction, with $\omega$, $\Sigma^+$, and $\pi^0$ as internal particles, and $\Sigma(1385)3/2^+$, $\Sigma(1380)1/2^-$ coupling to the final state $\pi^+\Lambda$. It is found that, instead of a usual Breit-Wigner peak, the existence of $\Sigma(1380)1/2^-$ is featured by a pronounced peak from the triangle singularity at $M_{\pi\Lambda}=1.35$ GeV, due to the $S$-wave couplings of $\Sigma(1380)1/2^-$ to $\Sigma^+\pi^0$ and $\pi^+\Lambda$. When $\Sigma(1380)1/2^-$ is absent, that peak is almost completely suppressed since the corresponding couplings of $\Sigma(1385)3/2^+$ are in $P$ wave, driving down the near-threshold reaction amplitudes. The uncertainties of this criterion may stem from the interference between the triangle loop and the tree-level diagrams, and the possible additional tree-level diagram of the radiative decay $\Lambda^+_c \to \gamma \Sigma^*$. It is shown that those cannot overturn our main conclusion, and can be clarified in a quantitative manner when the experimental data is available. As a side product, we also predict the branching ratio $\text{Br} \left( \Lambda^+_c \to \gamma \pi^+ \Lambda \right) \simeq 1.5 \times 10^{-5}$ assuming the existence of $\Sigma(1380)1/2^-$. We suggest the high-precision measurements of the $\Lambda^+_c \to \gamma \pi^+ \Lambda$ decay around $M_{\pi\Lambda}=1.35$ GeV by future experiments such as BESIII, LHCb, and Belle. 

\begin{acknowledgements}
This work is supported by the National Natural Science Foundation of China under Grants No.~12347155 and No.~12175240, the China Postdoctoral Science Foundation under Grant No.~2024M753172, and the Fundamental Research Funds for the Central Universities.
\end{acknowledgements}

\end{document}